# Web Service Testing Tools: A Comparative Study

**Shariq Hussain, Zhaoshun Wang, Ibrahima Kalil Toure and Abdoulaye Diop**

**School of Computer and Communication Engineering, University of Science and Technology Beijing**
**Beijing, 100083, China**
*s.hussain@zoho.com, zhswang@sohu.com, ikalil@msn.com, b20100556@xs.ustb.edu.cn*

**Abstract**
Quality of Service (QoS) has gained more importance with the increase in usage and adoption of web services. In recent years, various tools and techniques developed for measurement and evaluation of QoS of web services. There are commercial as well as open-source tools available today which are being used for monitoring and testing QoS for web services. These tools facilitate in QoS measurement and analysis and are helpful in evaluation of service performance in real-time network. In this paper, we describe three popular open-source tools and compare them in terms of features, usability, performance, and software requirements. Results of the comparison will help in adoption and usage of these tools, and also promote development and usage of open-source web service testing tools.

## 1. Introduction

The success of web service technology is clearly evident from the usage and adoption of this IT technology. A large number of providers from different sectors of industry are shifting to web service technology. Web services are software components accessible through programmatic interfaces and can perform tasks from simple requests to complex processes [1]. The heterogeneous nature of web service technology offers advantages like interoperability, usability, use of standardized communication protocol, deployability, etc. This makes web services technology an ideal candidate for organizations to host and deploy services in order to collaborate with other organizations in a flexible manner.

In order to attain the trust of service users, it is necessary that the system must conform to the performance requirements as it is the most important criteria for evaluating a system. It is therefore necessary to test the system before deployment in order to ensure that the system meets quality of service requirements. Various testing tools have been developed and designed for testing of web services. By using these test tools, web engineers can perform their tasks easily and efficiently, thus improving the quality of the system.

There are commercial as well as open-source test tools available in the market with different features and functionalities. In our study we are focusing on testing of Simple Object Access Protocol (SOAP) [2] web services. SOAP web services use XML language for definition of message architecture and message format. Web Services Description Language (WSDL) [3], an XML language is used to describe operations and interfaces of the web service. HTTP protocol is used for communication due to its wide usage and popularity.

Test tools automate the process of testing and are targeted to a specific test environment such as functional testing, performance testing, load testing, exception testing, etc. With the help of test tools, testers can create, manage and execute tests for a specific test environment for a particular application. The test results are compared with the expected results to evaluate the quality of the product.

Web service testing is a quite challenging area for researchers. The importance of this can also be judged with the ongoing research in this field. Several methods and techniques proposed by researchers as well as development of testing tools. There are commercial as well as open-source test tools available today for testing of web services.

Several studies are available which have compared various web service testing tools from functionalities, features, services, popularity, and so on. To our knowledge, there is still no comparative study on the representative testing tools, such as JMeter, soapUI, and Storm. In this paper, we compare these tools in terms of features, architecture, test environments, software requirements, and provide some observations. The comparison may help in selection of most suitable web service testing tool and promote the development and usage of open-source test tools.

This paper is organized as follows: Section 2 presents an overview of testing tools. Section 3 describes the three selected tools and their comparisons are reported in Section 4. Section 5 introduces related work and Section 6 concludes the paper.

## 2. Testing Tools

Software testing is the process of executing a program to verify its functionality and correctness [4]. Software testing is mostly deployed by programmers and testers. The aim of testing is to find the problems and to fix them to improve the software quality. Software testing methodology can be divided into two groups. One is manual testing and the other is automated testing. Manual Testing is a process in which testing process are carried out manually by the tester, usually follow a test plan comprised of test cases. On the other hand, automated testing is done with the help of automated test tools. Automated testing uses scripts to test operations of application automatically, reduces the need of human involvement and requires less time.

Software testing tools provide enormous aids to the testers in performing their tasks. Although the scope of testing tools is limited to particular test environments, the advantages associated are quite impressive. Benefits of automated testing includes: a better test coverage, quality improvements and more tests can be completed within a shorter time [5]. Tests can be performed to analyze the behavior of application in repeated executions of same operations. Further, testing tools perform testing faster than human users.

## 3. Overview of Open-Source Web Service Testing Tools

There is a number of open-source web service testing tools available in the software market. Although the core functions of these tools are similar, they differ in functionality, features, usability and interoperability. Keeping in view the above-mentioned aspects, we have selected three representative web service testing tools for comparison. Among them are, JMeter and soapUI are implemented in Java, and Storm is implemented in F# (F Sharp). A brief description of each of them is presented below.

### 3.1 JMeter

JMeter [6] is an open-source testing tool developed by Apache Software Foundation (ASF). It is distributed under Apache License. It was originally designed to test Web applications but has been extended to other test functions. The core function of JMeter is to load test client/server application but it can also be used for performance measurement. Further, JMeter is also helpful in regression testing by facilitating in creation of test scripts with assertions. By this way, we can verify that the application returns the expected results.

JMeter supports full multithreading that allows concurrent sampling by many threads and simultaneous sampling of different functions by separate thread groups. JMeter offers high extensibility due to use of pluggable components. These pluggable components include timers, samplers and visualization plugins. JMeter offers user-friendly Graphical User Interface (GUI). Configuration and setting up a test plan requires very little efforts. JMeter offers a number of statistical reports as well as graphical analysis. The latest release is version 2.8.

### 3.2 soapUI

soapUI [7] is an open-source testing tool for Service Oriented Architecture (SOA) [8] and web service testing. It is developed by SmartBear Software and is provided freely under the GNU LGPL. soapUI facilitates quick creation of advanced performance tests and execution of automated functional tests. The set of features offered by soapUI helps in performance evaluation of web services. Analysis of the test results provides a mean to improve the quality of services and applications.

soapUI offers easy-to-use GUI and is capable of performing variety of tests by offering many enterprise-class features. soapUI received a number of awards: ATI Automation Honors, 2009 [9], InfoWorld Best of Open Source Software Award, 2008 [10] and SOAWorld Readers' Choice Award, 2007 [11]. The latest version of soapUI is 4.5.1.

### 3.3 Storm

Storm [12] is a free and open-source tool for testing web services. It is developed by Erik Araojo. Storm is developed in F# language and is available for free to use, distributed under New BSD license.

Storm allows to test web services written using any technology (.Net, Java, etc.). Storm supports dynamic invocation of web service methods even those that have input parameters of complex data types and also facilitates editing/manipulation of raw soap requests. The GUI is very simple and user friendly. Multiple web services can be tested simultaneously that saves time, speed up testing schedule. Current stable version is r1.1-Adarna.

## 4. Comparison of Web Service Testing Tools

In this section, we present a comparison of the three web service testing tools, and then provide our observations. Such a comparison is helpful for the users/researchers to choose the suitable test tool for their needs.

Table 1: Technical overview of web service testing tools

| Tool | Technology Support | First release | Latest version/ Release date | Programming language | Operating System Support | Requirement | License | Developed by | Website |
|---|---|---|---|---|---|---|---|---|---|
| JMeter | Web-HTTP, HTTPS SOAP Database via JDBC LDAP JMS Mail-SMTP(S), POP3(S) and IMAP(S) Native commands or shell scripts | 2001 | 2.8 / Oct 6, 2012 | Java | Cross-platform | JRE 1.5+ | Apache License 2.0 | Apache Software Foundation | http://jmeter.apache.org/ |
| soapUI | Web-HTTP, HTTPS SOAP Database via JDBC JMS REST AMF | 2005 | 4.5.1 / Jun 27, 2012 | Java | Cross-platform | JRE 1.6+ | GNU LGPL 2.1 | SmartBear Software | http://www.soapui.org/ |
| Storm | SOAP | 2008 | 1.1 / Oct 29, 2008 | F# | Microsoft Windows | .NET Framework 2.0 F# 1.9.3.14 (optional) | New BSD License | Erik Araojo | http://storm.codeplex.com |

## 4.1 Technical Overview

The three testing tools chosen for comparison are based on different platforms and technologies. A detailed technical overview of them is shown in Table 1.

## 4.2 Comparison and Evaluation Approach

In order to compare the representative testing tools, we consider sample of three web services. The detail of web services is presented in Table 2.

To test the representative testing tools, each tool need to be configured to run the tests. The configuration includes installation, setting up test environment, test parameters, test data collection, reports analysis, etc. Each tool is configured to test the sample web services and gather test results.

We run the tests on an Intel Core 2 Duo 2.0 GHz processor machine with 3GB RAM, running Microsoft Windows 7 Ultimate, and 2Mbps of DSL Internet connection.

The tests were conducted four times a day at regular intervals to get fair and transparent results. The reason was to minimize the affect of Internet connection's performance on the test results and to obtain realistic measurements. The performance of Internet varies depending on the time of day and other factors such as internet traffic, subscribed users, etc.

Table 2: Sample web services

| ID | Web Service Name | Description | Publisher |
|---|---|---|---|
| W1 | TempConvert | Conversions from Celsius to Fahrenheit and vice versa | W3Schools |
| W2 | Weather | Allows to get city's weather | CDYNE Corporation |
| W3 | ZipCode | Returns a list of City+State for a supplied zip code | Ripe Development LLC |

Table 3: Minimum and maximum response time of testing tools for web services

| Tool | Web Service ID | Response Time (ms) | | | | | | | |
|---|---|---|---|---|---|---|---|---|---|
| | | 12:00 AM | | 6:00 AM | | 12:00 PM | | 6:00 PM | |
| | | Min | Max | Min | Max | Min | Max | Min | Max |
| JMeter | W1 | 1237 | 4906 | 1056 | 4304 | 1077 | 1921 | 1147 | 4320 |
| | W2 | 1880 | 18276 | 1121 | 16087 | 1595 | 19056 | 1523 | 18984 |
| | W3 | 954 | 25660 | 806 | 3852 | 866 | 10023 | 912 | 7052 |
| soapUI | W1 | 334 | 1423 | 300 | 1158 | 307 | 1424 | 299 | 4048 |
| | W2 | 557 | 60011 | 315 | 12062 | 402 | 6124 | 527 | 16096 |
| | W3 | 639 | 7113 | 534 | 6750 | 576 | 9761 | 625 | 7002 |
| Storm | W1 | 666 | 3581 | 577 | 1482 | 593 | 1298 | 624 | 1794 |
| | W2 | 1060 | 15179 | 619 | 99013 | 718 | 7318 | 936 | 32417 |
| | W3 | 998 | 7634 | 822 | 2246 | 852 | 6895 | 936 | 4103 |

The three selected tools were tested by invoking the sample web services for a pre-defined sample count. The results were collected and compiled for analysis.

4.3 Results and Discussions

In this section we describe different comparative results with testing tools.

Each tool has different architecture and internal processes to carry out tasks. This factor provides basis to compare the tools in terms of response time. Minimum and maximum values for response time at different time intervals are shown in Table 3.

From Table 3 we observe that the response time values taken at 6:00 AM are most optimal. This shows that the performance of Internet connection is better at 6:00 AM which is reflected in response time values.

Further, results of the tests are summarized to calculate average response time of each test tool for each web service. Table 4 shows the average response time of each tool for each web service. This data is also presented in the form of graph as shown in Figure 1.

Table 4: Average response time of testing tools

| ID | Average Response Time (ms) | | |
|---|---|---|---|
| | JMeter | soapUI | Storm |
| W1 | 1359.83 | 401.44 | 758.73 |
| W2 | 3541.33 | 1193.01 | 1939.92 |
| W3 | 1357.25 | 1046.78 | 1350.33 |

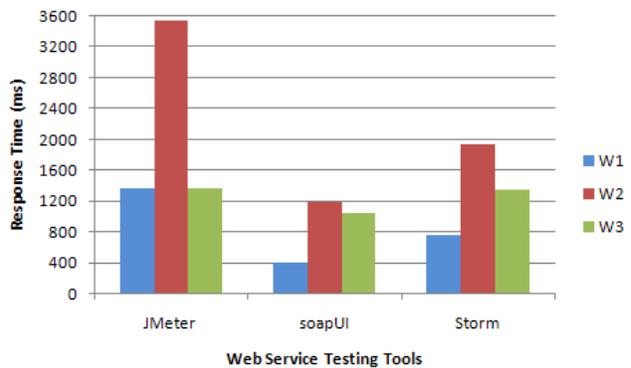

Fig. 1 Average response time of testing tools for sample web services.

From the results, we observe that JMeter is taking more time in responding to web services as compared to other two tools. Storm is behaving better than JMeter but not promising as soapUI. In this test, soapUI outperforms other two testing tools and can be regarded as fastest tool in terms of response time.

The next comparison is based on the average throughput criterion. Throughput is the measure of the number of requests that can be served by web service in a specified time period [13]. Only JMeter and soapUI supports this type of testing and provides information on throughput test results. Average throughput of each tool for each web service is shown in Figure 2.

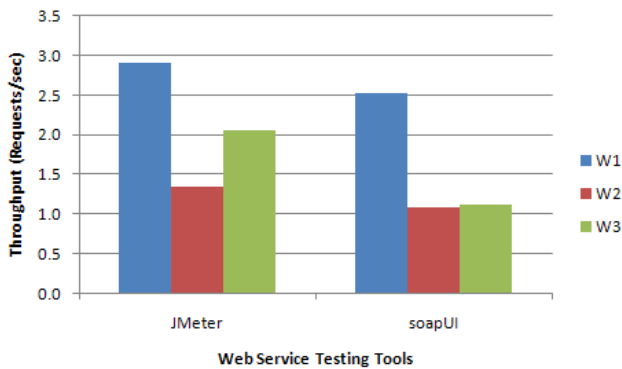

Fig. 2  Average throughput of testing tools for sample web services.

The results of throughput test demonstrate that JMeter has better throughput than soapUI. In case of W3, JMeter shows more than 84% throughput than soapUI, while for W1 and W2, increase is 14.5% and 24% respectively. Therefore, JMeter has better throughput than soapUI.

Another parameter that is observed during testing of web services is number of bytes per second processed by the test. Figure 3 shows the KB/sec values of both tools.

From Figure 3 it is seen that number of bytes processed by JMeter is higher than soapUI. This is in relation to the throughput attribute, as JMeter has better throughput as shown in Figure 2. This means that JMeter has processed more bytes during test as its throughput is better than other tool.

Further useful information related to testing, reported by JMeter and soapUI contains number of assertion errors and number of lost samples (failed request ratio).

Usability is another factor for evaluation of testing tools. The study shows that the Storm has a very simple and easy to use interface. Simply by adding WSDL reference, supported methods are displayed for invocation. To perform a test one need to provide input parameters for methods. The result of the web service call along with SOAP response and elapsed time for the test is returned to user. soapUI has attractive graphical user interface with many useful test utilities. Steps to perform a test includes, creation of a new project with name and WSDL reference, addition of web service methods into new request and finally creation of load test. LoadTest window provides statistical information about test along with graphical representation. JMeter has an excellent user interface with an iconic toolbar and a right pane to display the details of each element of test plan. For web service testing, a new test plan has to be created with thread group, loop controller, timer and sampler. Each element need to be placed in a parent child relationship in a hierarchical form. Presently, JMeter doesn't support automatic WSDL handling. There are three options for the post data (soap message): text area, external file, or directory. Different types of listeners are available to show test results in various forms. JMeter supports a lot of different types of test plans.

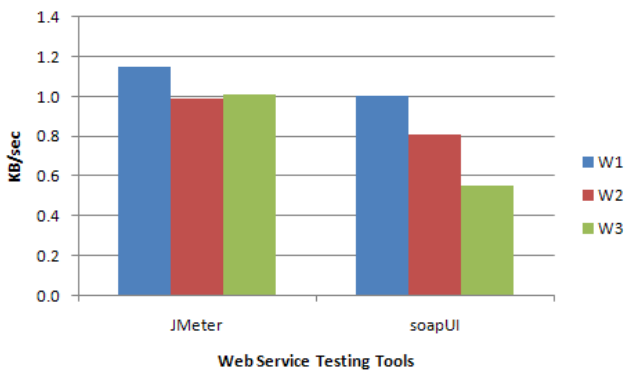

Fig. 3  Average Kilobytes per second of testing tools.

Comparison of different testing tools is a complex task due to the fact that testing tools may not comply with same test criteria i.e. one tool may have the ability to test throughput (in our case JMeter and soapUI), while another tool i.e. Storm, does not have this criteria. Furthermore, one tool may have better performance in one test case, while poorer in other test criteria. For example, in our study soapUI has better response time but throughput is not as good as JMeter's throughput.

## 5. Related Work

Since the beginning of web service testing, different approaches have been proposed in literature. In this section, we describe several closely related work.

Performance testing of web services using JMeter is demonstrated in great detail [14]. JMeter is also able to perform load testing of web applications especially J2EE-based web applications [15-16]. In [17], authors proposed test steps of the web service testing tools for testing an orchestration and showed the applicability with soapUI. In some approaches, testing of web services based on WSDL descriptions, soapUI tool is used for derivation of SOAP envelope skeleton [18-19]. soapUI is also used for testing of sample web services developed using SAS BI platform [20]. The study [21] presented a preliminary approach towards an evaluation framework for SOA security testing tools and tested soapUI for suitability assessment. A research study on different test tools and techniques for testing atomic and composed services is presented along with development of a prototype for automated choreography tests [22]. Security testing is also a challenging task in SOA domain. Altaani and Jaradat [23] analyzed the security requirements in SOA and presented techniques for security testing, validated the results by using soapUI. A comparative study of three web service testing tools with several selected web services is done in which soapUI outperforms other two tools [24].

In the paper, we presented an overview of three open-source web service testing tools and a technical overview of each tool. We also provide a comprehensive comparison of them, which may help researchers in selection of suitable tool.

## 6. Conclusions

Nowadays we can see that web service technology turn out to be the latest trend and provides a new model of web. The rapid growth of web service market necessitated developing of testing methodologies, hence different methods and different tools proposed to test web services. In this paper, we present a comparative study of open-source web service testing tools with technical overview and features. Comparison is made on several quality factors including response time, throughput, and usability. Tools are evaluated by collecting the sample web services and collecting the test results. The comparison may give researchers an informative overview with potential benefits of open-source testing tools, and also help in promotion and development of open-source testing tools.

**Acknowledgments**

The work reported in this paper was supported by the National Natural Science Foundation of China (Grant No. 60903003), the Beijing Natural Science Foundation of China (Grant No. 4112037), and the Research Fund for the Doctoral Program of Higher Education of China (Grant No. 2008000401051).

**Shariq Hussain** received his Master's degree in Computer Science from PMAS Arid Agriculture University, Rawalpindi, Pakistan, in 2007. He is now a PhD student in the School of Computer and Communication Engineering, University of Science and Technology Beijing. His main research interests include web service QoS, web service monitoring, web service testing and e-learning.

**Zhaoshun Wang** is a Professor and the Associate Head of the Department of Computer Science of the University of Science and Technology Beijing. He graduated from Department of Mathematics, Beijing Normal University in 1993. He received his PhD from Beijing University of Science and Technology in 2002. He completed postdoctoral research work at the Graduate School of the Chinese Academy of Sciences from 2003 to 2006. Teaching and research work, research direction has been engaged in the direction of computer software for software engineering, software security, information security, ASIC chip design. In recent years, participated in the national "863", "973", the National Natural Science Foundation of China, National "Eleventh Five-Year" password Fund, Outstanding Young Teachers Fund of the Ministry of Education, Beijing natural science fund of each one; auspices of the National Information Security Standardization Technical Committee Project 2, and 8 Plant Association project topics. He has published more than 60 scientific papers in core computer science journals and international conferences, which retrieve an SCI, EI retrieval of more than 10 articles, ISTP retrieval of more than 10 articles; teaching and research of more than 10 papers. Further achievements include a provincial and municipal Science and Technology Progress Award, two national invention patents, two textbooks, and a monograph.

**Ibrahima Kalil Toure** received his Master's degree in Programming Analysis from University of Conakry, Conakry, Guinea, in 2000. He is now a PhD student in the School of Computer and Communication Engineering, University of Science and Technology Beijing. His main research interests include web services and composition.

**Abdoulaye Diop** is currently a PhD student in the School of Computer and Communication Engineering, University of Science and Technology Beijing. His main research interests include wireless sensor networks and security.